\documentclass[journal]{IEEEtran}
\usepackage[T1]{fontenc}
\usepackage{cite}
\usepackage{amsmath,amssymb}
\usepackage{graphicx}
\usepackage{booktabs}
\usepackage{multirow}
\usepackage{tikz}
\usetikzlibrary{positioning,arrows.meta,fit,patterns,calc,backgrounds}
\usepackage{pgfplots}
\pgfplotsset{compat=1.17}
\usepackage{algorithm}
\usepackage{algpseudocode}
\usepackage{xspace}
\usepackage{url}
\newcommand{\papertitle}{WEPWAWET-CORE: A Hardware-Resident ML-DSA Architecture with Fabric-Generated, Fabric-Sealed Keys for Post-Quantum Avionics}
\newcommand{\wc}{WEPWAWET-CORE\xspace}
\newcommand{\rhopp}{\rho''}

\begin{document}
\title{\papertitle}
\author{Mohamed~El-Hadedy,~\IEEEmembership{Senior~Member,~IEEE,}
        and~Wen-Mei~Hwu,~\IEEEmembership{Fellow,~IEEE}
\thanks{This work has been submitted to the IEEE for possible publication. Copyright may be transferred without notice, after which this version may no longer be accessible.}%
\thanks{M. El-Hadedy is with the Department of Electrical and Computer
Engineering, California State Polytechnic University, Pomona, CA, USA,
and also with the Coordinated Science Laboratory, University of Illinois
at Urbana-Champaign, Urbana, IL, USA (e-mail: mealy@cpp.edu;
hadedy@illinois.edu). (Corresponding author: Mohamed El-Hadedy.)}
\thanks{W.-M. Hwu is with the Coordinated Science Laboratory, University
of Illinois at Urbana-Champaign, Urbana, IL, USA (e-mail: w-hwu@illinois.edu).}}

\maketitle
\begin{abstract}
Spacecraft that exchange commands over crosslinks need quantum-resistant signatures, for which the Module-Lattice-Based Digital Signature Algorithm (ML-DSA, FIPS~204) is the standard, yet a software signer runs its data-dependent rejection loop, and keeps its key, on cores and memory shared with attitude control and payload processing. This paper shows that for a hardware-assisted signer the cost that matters is not the arithmetic but the number of times the processor crosses into the fabric, and that the boundary which minimizes those crossings is also where the key can be sealed. A gem5-supported analysis of three control granularities over the same kernels shows that only one job per signature, with the whole rejection loop on chip, takes the processor off the critical path. \wc realizes that boundary on an XCZU3EG at 300~MHz: a ROM microsequencer runs Sign, Verify, and key generation as three programs over one datapath and a banked scratchpad, and keeps the key, the on-die entropy, and the mask seed off the processor bus. The processor-side cost of a signature falls from over a second under per-kernel control to 0.53~ms, fixed whatever the number of rejections, and the fabric's time per attempt varies by less than 0.1\% under a saturating workload. Under NASA JPL's F$'$ with attitude-control and imaging loads, two nodes with fabric-generated, ground-certified keys exchanged 3{,}575 signed frames, every frame was accepted, and no attitude deadline was missed. A cycle profile places the remaining distance to dedicated cores in two serial units.
\end{abstract}

\section{Introduction}
\IEEEPARstart{A}{} swarm of small satellites that shares attitude and state estimates over crosslinks works only if each node can authenticate what its neighbors send, for the length of the mission. The pairwise authenticated link is the element such a swarm is built from, and it is what this paper builds and measures between two nodes. Missions fielded today will still be flying when cryptographically relevant quantum computers may exist~\cite{mosca2015}. NIST standardized ML-DSA as a post-quantum signature in FIPS~204~\cite{fips204}, and the NSA's Commercial National Security Algorithm (CNSA) 2.0 suite requires the ML-DSA-87 parameter set for national security systems, with new acquisitions expected to comply from January 2027~\cite{cnsa2}, so signing crosslink commands with ML-DSA is a near-term requirement for space systems in that category.

Raw speed is not the obstacle on a flight computer. On the Cortex-A53 cores of a Zynq UltraScale+ device the PQClean implementation for the Arm Neon vector extension~\cite{kannwischer2022pqclean,becker2022neon} signs ML-DSA-44 in 0.89~ms at the median.
The obstacle is where that work runs. Signing executes a rejection loop whose iteration count is random, so its tail is several times its median, and the loop shares cores, caches, and DRAM with attitude estimation, payload compression, and the flight-software framework; the key and every per-signature seed also live in the operating system's memory.

Hardware for Dilithium and ML-DSA is mature as a core. Dedicated designs sign in about a hundred microseconds~\cite{beckwith2021,norga2026mldsaosh}, and instruction-driven lattice processors run key generation, signing with its rejection loop, and verification from one datapath~\cite{aikata2021unified,aikata2023kali}. These designs are evaluated at the core boundary, by cycles and resources, and Section~\ref{sec:related} examines what they leave open: where the randomness comes from, whether the host can read the key, and what the signer costs inside a running flight computer.

\wc follows the programmable line for FIPS~204 ML-DSA-44 and starts from an observation about cost: once the kernels sit in the fabric, what a signature costs the processor is the number of times it crosses into the fabric and not the arithmetic, and the boundary that minimizes those crossings, one job per signature, is also the boundary behind which the key can be sealed. Four questions follow from that observation and organize the paper. Where does a hardware-assisted signature spend its time, in arithmetic or in orchestration (Section~\ref{sec:analysis})? Can a small FPGA own the complete control of key generation, signing with its rejection loop, and verification on one datapath (Sections~\ref{sec:arch} and~\ref{sec:impl})? What does such an engine give and cost inside running flight software (Sections~\ref{sec:system} and~\ref{sec:mission})? And does the boundary keep the key, the randomness, and the mask seed out of the processor's reach (Section~\ref{sec:security})? The answers are an analysis that derives the architecture from the count of bus accesses; a ROM microsequencer, a banked scratchpad, and a retry manager that run Sign, Verify, and key generation as three programs over one datapath at 300~MHz on an XCZU3EG whose slices are 90\% occupied; an entropy source, its conditioning, and the hedged seed derivation placed in the fabric with the key, an isolation that is attacked from the host on the board; and a measurement of two such nodes under F$'$ with an attitude-control and imaging workload. The design gives up the throughput of a dedicated pipeline for that generality and isolation, and Section~\ref{sec:impl} locates the difference in two serial units and prices its removal.

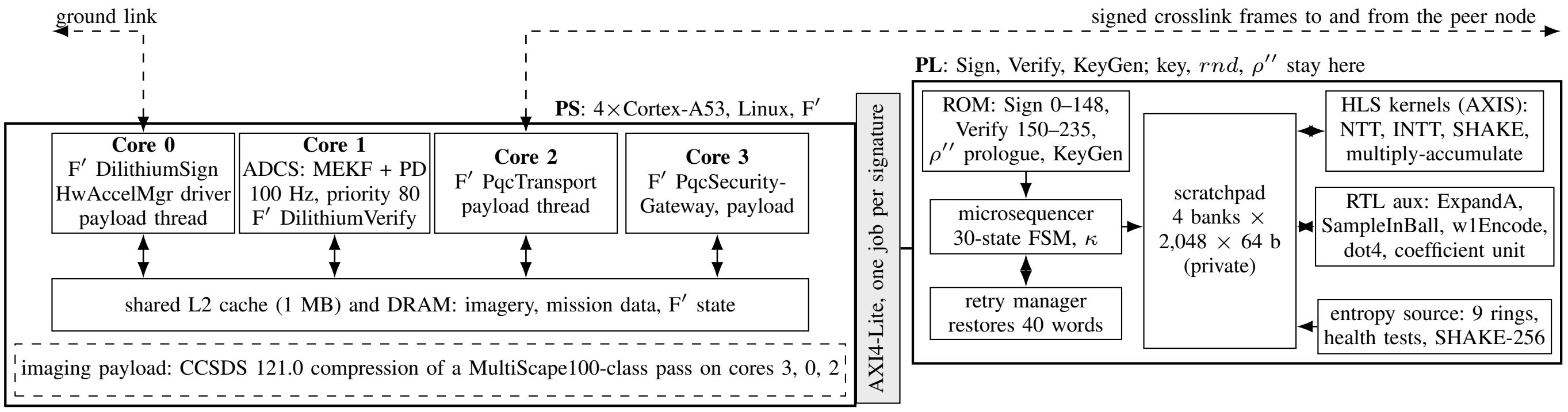
\begin{figure*}[t]
\centering
\begin{tikzpicture}[font=\scriptsize, >=Latex,
  core/.style={draw, minimum width=2.1cm, minimum height=1.15cm, align=center, inner sep=1pt},
  blk/.style={draw, align=center, inner sep=2pt, minimum height=6mm},
  arr/.style={->, line width=0.5pt}, dbl/.style={<->, line width=0.5pt}]
\node[core] (c0) at (0,0.45)   {\textbf{Core 0}\\F$'$ DilithiumSign\\HwAccelMgr driver\\payload thread};
\node[core] (c1) at (2.2,0.45) {\textbf{Core 1}\\ADCS: MEKF + PD\\100~Hz, priority 80\\F$'$ DilithiumVerify};
\node[core] (c2) at (4.4,0.45) {\textbf{Core 2}\\F$'$ PqcTransport\\payload thread};
\node[core] (c3) at (6.6,0.45) {\textbf{Core 3}\\F$'$ PqcSecurity-\\Gateway, payload};
\node[blk, minimum width=8.7cm] (l2) at (3.3,-0.95) {shared L2 cache (1~MB) and DRAM: imagery, mission data, F$'$ state};
\node[blk, minimum width=8.7cm, dashed] (pay) at (3.3,-1.7) {imaging payload: CCSDS 121.0 compression of a MultiScape100-class pass on cores 3, 0, 2};
\foreach \c in {c0,c1,c2,c3} \draw[dbl] (\c.south) -- (\c.south|-l2.north);
\node[draw, thick, fit=(c0)(c3)(l2)(pay), inner sep=3pt] (ps) {};
\node[anchor=south east, inner sep=1pt] at ([xshift=-10pt]ps.north east) {\textbf{PS}: 4$\times$Cortex-A53, Linux, F$'$};
\node[draw, fill=black!8, minimum width=3.0cm, minimum height=0.5cm, align=center, rotate=90] (axi) at (8.45,-0.3) {AXI4-Lite, one job per signature};
\node[blk, minimum width=2.2cm] (rom) at (10.15,1.05) {ROM: Sign 0--148,\\Verify 150--235,\\$\rhopp$ prologue, KeyGen};
\node[blk, minimum width=2.2cm] (seq) at (10.15,-0.05) {microsequencer\\30-state FSM, $\kappa$};
\node[blk, minimum width=2.2cm] (ret) at (10.15,-1.05) {retry manager\\restores 40 words};
\node[blk, minimum width=1.75cm, minimum height=2.7cm] (sp) at (12.38,-0.1) {scratchpad\\4 banks $\times$\\2{,}048 $\times$ 64~b\\(private)};
\node[blk, minimum width=2.5cm] (krn) at (14.85,1.05) {HLS kernels (AXIS):\\NTT, INTT, SHAKE,\\multiply-accumulate};
\node[blk, minimum width=2.5cm] (aux) at (14.85,-0.05) {RTL aux: ExpandA,\\SampleInBall, w1Encode,\\dot4, coefficient unit};
\node[blk, minimum width=2.5cm] (trng) at (14.85,-1.2) {entropy source: 9 rings,\\health tests, SHAKE-256};
\draw[arr] (rom) -- (seq); \draw[dbl] (seq) -- (ret); \draw[arr] (seq.east) -- (sp.west|-seq);
\draw[dbl] (sp.east|-krn) -- (krn.west); \draw[dbl] (sp.east|-aux) -- (aux.west); \draw[arr] (trng.west) -- (sp.east|-trng);
\node[draw, thick, fit=(rom)(ret)(sp)(krn)(trng), inner sep=3pt] (pl) {};
\node[anchor=south west, inner sep=1pt] at (pl.north west) {\textbf{PL}: Sign, Verify, KeyGen; key, $rnd$, $\rhopp$ stay here};
\draw[thick] (ps.east|-axi) -- (axi.north); \draw[thick] (axi.south) -- (pl.west|-axi);
\draw[dbl, dashed] (c2.north) -- (4.4,2.2) -- (16.3,2.2) node[pos=0.98, above left, inner sep=1pt]{signed crosslink frames to and from the peer node};
\draw[dbl, dashed] (c0.north) -- (0,2.2) -- (-1.05,2.2) node[pos=1, above right, inner sep=1pt]{ground link};
\end{tikzpicture}
\caption{A \wc spacecraft node.}
\label{fig:node}
\end{figure*}

Fig.~\ref{fig:node} shows the resulting node, with its processing system (PS) and its programmable logic (PL), the fabric. Section~\ref{sec:related} reviews related work and Section~\ref{sec:prelim} ML-DSA; the remaining sections take the four questions in order.

\section{Related Work}\label{sec:related}
Dedicated cores set the speed reference. The pipelined core of Beckwith \emph{et al.} restarts on rejection and signs Dilithium-2 in 117~$\mu$s on average~\cite{beckwith2021}, ML-DSA-OSH reaches 106~$\mu$s for ML-DSA-44~\cite{norga2026mldsaosh}, a compact design trades speed for area~\cite{land2021}, and low-complexity multipliers reduce the arithmetic of lightweight lattice schemes~\cite{he2021jetcas}. In these cores the seed is an input or its source is not stated.

The closest family is the instruction-driven lattice processor. The Saber and Dilithium cryptoprocessor of Aikata \emph{et al.} loads its program into an instruction memory and closes the rejection loop with a jump of the instruction pointer~\cite{aikata2021unified}; KaLi continues that line for Kyber and Dilithium with an 18-instruction set~\cite{aikata2023kali}; Sapphire keeps an instruction memory in the core but leaves the order of its Dilithium programs to RISC-V software~\cite{banerjee2019sapphire}; and Adams Bridge sequences ML-DSA-87 for a root of trust, with seed and randomness supplied by firmware or a key vault~\cite{bisheh2026adams}. \wc follows this line for ML-DSA-44 with its three programs resident in ROM, so the programmed shared datapath is not what it contributes. It differs in that the entropy source, its conditioning, and the hedged seed derivation sit in the fabric, and in that its latency is a measured mean over rejections where the Aikata designs report the best case.

Co-designs keep the algorithm flow on a processor and accelerate its kernels~\cite{mao2023codesign}, and the instruction-set extension HORCRUX attaches them to a RISC-V core as a 2{,}820-LUT coprocessor while the flow stays in software~\cite{dolmeta2025horcrux}. PQC-HA maps each Kyber and Dilithium operation to one HLS block, measures about 10~$\mu$s of communication overhead per call over PCIe, and argues that an off-chip accelerator should unite most of the function to avoid frequent transfers~\cite{sattel2023pqcha}. These works fix one division of control; Section~\ref{sec:analysis} varies it over the same kernels and separates the cost of host transactions from the cost of the arithmetic.

On-device key custody has been reported for ML-DSA-65 by Park \emph{et al.}~\cite{park2026aipqc}, whose entropy combines an on-die ring-oscillator source with off-chip radioactive-decay sources, and Adams Bridge hides the key from software in its key-vault mode; in the designs examined that isolation is stated rather than attacked from the host. In aerospace, ML-DSA has been measured as software in an avionics partition on a desktop processor~\cite{varner2025avionics} and in link simulation~\cite{kwon2026rtt}. No evaluation of an ML-DSA hardware signer inside flight software under a concurrent workload was found other than the signing service of~\cite{elhadedy2026wepwawet}, which kept the control flow on the processor and needed 1{,}295~ms per signature through the flight-software stack. Table~\ref{tab:feat} places the designs side by side.

\begin{table}[t]
\caption{\wc and prior lattice signature hardware.}\label{tab:feat}
\centering\footnotesize\setlength{\tabcolsep}{3.5pt}
\begin{tabular}{lccccc}
\toprule
Design & K/S/V & Loop & Entropy & Sealed & In system\\
\midrule
Beckwith~\cite{beckwith2021} & yes & yes & no & n.r. & no\\
Land~\cite{land2021} & yes & yes & n.r. & n.r. & no\\
ML-DSA-OSH~\cite{norga2026mldsaosh} & yes & yes & n.r. & n.r. & no\\
Aikata~\cite{aikata2021unified} & yes & yes & n.r. & n.r. & no\\
KaLi~\cite{aikata2023kali} & yes & n.r. & n.r. & n.r. & no\\
Sapphire~\cite{banerjee2019sapphire} & yes & host & off chip & n.r. & no\\
HORCRUX~\cite{dolmeta2025horcrux} & kernels & host & n.r. & n.r. & no\\
Adams Bridge~\cite{bisheh2026adams} & yes & yes & no & mode & no\\
Park~\cite{park2026aipqc} & yes & yes & hybrid & mode & no\\
\textbf{\wc} & yes & yes & yes & yes & yes\\
\bottomrule
\end{tabular}\\[2pt]
\raggedright\scriptsize K/S/V: key generation, signing, and verification all in hardware; kernels: only the kernels. Loop: rejection loop controlled in hardware. Entropy: source on the signer's chip; hybrid: combined with off-chip sources. Sealed: secret key unreadable over the host bus; mode: in one operating mode only. In system: evaluated inside a running system under load. n.r.: not reported in the text examined.
\end{table}

\section{Preliminaries}\label{sec:prelim}
ML-DSA uses SHAKE-128 and SHAKE-256~\cite{fips202} for all expansion and hashing and works over $R_q=\mathbb{Z}_q[X]/(X^{256}+1)$ with $q=8{,}380{,}417$~\cite{fips204}. ML-DSA-44 uses a $4\times4$ public matrix $\mathbf{A}$ expanded from a 32-byte seed $\rho$, secret vectors $\mathbf{s}_1,\mathbf{s}_2$ with coefficients in $[-2,2]$, and the public $\mathbf{t}=\mathbf{A}\mathbf{s}_1+\mathbf{s}_2$, split by Power2Round into $\mathbf{t}_1$ (public) and $\mathbf{t}_0$ (secret). Products are formed in the domain of the number-theoretic transform (NTT), marked by a hat. Algorithm~\ref{alg:sign} shows the signing procedure as \wc executes it, with the hedged mask seed $\rhopp=H(K\,\|\,rnd\,\|\,\mu)$ of FIPS~204. In it $H$ is SHAKE-256, $K$ the 32-byte signing seed held in the secret key, $\mu$ the 64-byte digest of public key and message, $rnd$ a fresh 32-byte random string, $\kappa$ the mask counter, and $c$ the challenge polynomial; the bounds are $\gamma_1=2^{17}$, $\gamma_2=(q-1)/88$, $\beta=78$, and $\omega=80$, and Power2Round drops $d=13$ bits. One attempt forms the mask $\mathbf{y}$, the commitment $\mathbf{w}$, the challenge seed $\tilde c$, the response $\mathbf{z}$, and the hint $\mathbf{h}$, and $\sigma$ is the signature; $\circ$ is the coefficientwise product, $\|$ concatenation, $\|\cdot\|_\infty$ the largest coefficient magnitude, and $\|\mathbf{h}\|_1$ the number of ones in $\mathbf{h}$. The capitalized functions are those of FIPS~204.

\begin{algorithm}[t]
\caption{ML-DSA-44 Sign (hedged).}\label{alg:sign}
\small
\begin{algorithmic}[1]
\Require $\rho$, $\mathbf{s}_1,\mathbf{s}_2,\mathbf{t}_0$, $K$, $\mu$, $rnd$ from the on-die entropy source
\State $\rhopp\gets H(K\|rnd\|\mu)$
\State $\hat{\mathbf{A}}\gets\textsc{ExpandA}(\rho)$, $\kappa\gets0$
\Loop
 \State $\mathbf{y}\gets\textsc{ExpandMask}(\rhopp,\kappa)$; $\mathbf{w}\gets\mathrm{NTT}^{-1}(\hat{\mathbf{A}}\circ\mathrm{NTT}(\mathbf{y}))$
 \State $\tilde c\gets H(\mu\|\textsc{w1Encode}(\textsc{HighBits}(\mathbf{w})))$
 \State $c\gets\textsc{SampleInBall}(\tilde c)$
 \State $\mathbf{z}\gets\mathbf{y}+c\mathbf{s}_1$; $\mathbf{r}_0\gets\textsc{LowBits}(\mathbf{w}-c\mathbf{s}_2)$
 \If{$\|\mathbf{z}\|_\infty\ge\gamma_1-\beta$ \textbf{or} $\|\mathbf{r}_0\|_\infty\ge\gamma_2-\beta$} $\kappa\gets\kappa+4$; \textbf{continue}\EndIf
 \State $\mathbf{h}\gets\textsc{MakeHint}(-c\mathbf{t}_0,\mathbf{w}-c\mathbf{s}_2+c\mathbf{t}_0)$
 \If{$\|c\mathbf{t}_0\|_\infty\ge\gamma_2$ \textbf{or} $\|\mathbf{h}\|_1>\omega$} $\kappa\gets\kappa+4$; \textbf{continue}\EndIf
 \State \Return $\sigma=(\tilde c,\mathbf{z},\mathbf{h})$
\EndLoop
\end{algorithmic}
\end{algorithm}

The loop count of Algorithm~\ref{alg:sign} is random, and that randomness is the property any hardware realization has to plan around. FIPS~204 gives 4.25 expected repetitions for ML-DSA-44~\cite{fips204}; an instrumented PQClean build averages 4.37 over 80{,}000 signatures, and the fabric's own counter averaged 4.49 over 4{,}234 hedged signatures.

The mask seed $\rhopp$ is equivalent to a secret: it determines $\mathbf{y}$, and with $\mathbf{z}=\mathbf{y}+c\mathbf{s}_1$ published, an observer of $\mathbf{y}$ learns $\mathbf{s}_1$. Hedging with fresh $rnd$ protects against the fault attacks that exploit deterministic nonces~\cite{bruinderink2018dfa}, but only when $rnd$ is fresh and unpredictable, so the entropy source becomes part of the signer.

\begin{algorithm}[t]
\caption{ML-DSA-44 Verify.}\label{alg:verify}
\small
\begin{algorithmic}[1]
\Require $\rho$, $\mathbf{t}_1$, $\mu$, $\sigma=(\tilde c,\mathbf{z},\mathbf{h})$
\State $\hat{\mathbf{A}}\gets\textsc{ExpandA}(\rho)$; $c\gets\textsc{SampleInBall}(\tilde c)$
\State $\mathbf{w}'\gets\mathrm{NTT}^{-1}(\hat{\mathbf{A}}\circ\mathrm{NTT}(\mathbf{z})-\mathrm{NTT}(c)\circ\mathrm{NTT}(\mathbf{t}_1 2^d))$
\State $\mathbf{w}'_1\gets\textsc{UseHint}(\mathbf{h},\mathbf{w}')$
\State $ok\gets\|\mathbf{z}\|_\infty<\gamma_1-\beta$ \textbf{and} $\|\mathbf{h}\|_1\le\omega$
\State \Return $ok$ \textbf{and} $\tilde c=H(\mu\|\textsc{w1Encode}(\mathbf{w}'_1))$
\end{algorithmic}
\end{algorithm}

Algorithm~\ref{alg:verify} is verification. It needs only the public key, has no loop, and uses the same NTT, multiply-accumulate, hashing, and coefficient primitives as signing, which is the observation that lets one datapath serve both.

The system in which this signature runs has two spacecraft, each with a Zynq UltraScale+ flight computer whose four Cortex-A53 cores run Linux and the F$'$ flight-software framework~\cite{bocchino2018fprime}. Each node holds its own ML-DSA-44 key pair and the public key of its peer, and sends authenticated command frames to its peer. A frame is a text envelope naming the protocol version, sender, receiver, a sequence number, and the command (for example \texttt{WCORE-V1|AUP1|AUP2|SEQ=n|PeerPing}) followed by its ML-DSA signature; the receiver verifies the signature before the command reaches any consumer and rejects replays by sequence number.
The application cores also run the mission: an attitude determination and control system (ADCS) task at 100~Hz and an imaging payload whose frames are compressed with the Consultative Committee for Space Data Systems (CCSDS) 121.0 lossless algorithm~\cite{ccsds121} before downlink. The ADCS task is the hard real-time function of a small satellite, since a control period released late is a pointing error; its release delay is therefore the measure used here of whether signing disturbs the flight computer, and the payload is the throughput load that fills the remaining cores. Link-layer protection of the same frames is the subject of CCSDS Space Data Link Security~\cite{ccsds355}; this work concerns the signature engine such a protocol would call.

\section{Where Does the Time Go?}\label{sec:analysis}
\begin{figure}[t]
\centering
\begin{tikzpicture}
\begin{axis}[width=\columnwidth, height=3.4cm, ybar, bar width=13pt, ymode=log, ymin=0.5, ymax=4000,
  log origin=infty, symbolic x coords={percall,perstage,persig}, xtick=data,
  xticklabels={{Kernel-per-call\\(replay, measured\\access cost)},{Per-stage chaining\\(model)},{Per-signature\\(this work,\\measured)}},
  xticklabel style={align=center, font=\scriptsize}, ylabel={ms per signature}, ylabel style={font=\scriptsize},
  yticklabel style={font=\scriptsize}, enlarge x limits=0.22, axis x line*=bottom, axis y line*=left,
  nodes near coords, nodes near coords style={font=\scriptsize, above},
  point meta=explicit symbolic, clip=false]
\addplot[draw=black, pattern=north east lines] coordinates {(percall,1128)[$\ge$1{,}128] (perstage,800)[$\ge$800] (persig,7.48)[7.5]};
\draw[dashed, thick] ({rel axis cs:0,0}|-{axis cs:percall,1.10}) -- ({rel axis cs:1,0}|-{axis cs:percall,1.10}) node[pos=1, above left, font=\scriptsize, fill=white, inner sep=1pt]{Neon software, 1.1~ms};
\node[font=\scriptsize, align=left, anchor=west, xshift=8pt] at (axis cs:percall,120) {bus 119\\waits 1{,}009};
\node[font=\scriptsize, align=left, anchor=west, xshift=8pt] at (axis cs:perstage,90) {bus 36--45\\waits 761};
\end{axis}
\end{tikzpicture}
\caption{Time per signature for three control styles.}
\label{fig:dse}
\end{figure}
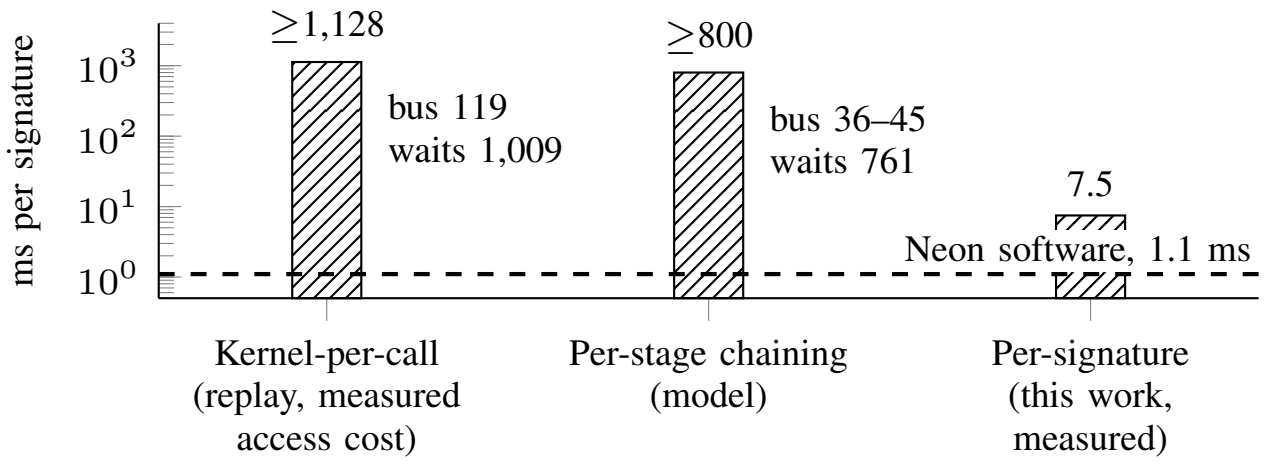

The first question is whether the arithmetic or its orchestration dominates a signature once the kernels sit in the fabric, and the answer decides the architecture. Consider one family of fabric kernels, forward and inverse NTT, SHAKE, and polynomial multiply-accumulate, driven at three control granularities: the processor issues every kernel call and moves every operand word (kernel-per-call), the processor chains the kernels of a stage so that intermediate results stay in the fabric (per-stage), or the fabric runs the whole rejection loop and the processor invokes it once per signature as a single job (per-signature). Fig.~\ref{fig:dse} gives the time each needs per mean signature.

\begin{figure}[t]
\centering
\begin{tikzpicture}
\begin{axis}[width=\columnwidth, height=3.4cm, xmode=log, ymode=log, xmin=40, xmax=1250, ymin=0.15, ymax=2000,
  xtick={50,200,1000}, xticklabels={50,200,1000}, xlabel={register access latency (ns)}, ylabel={ms per signature},
  xlabel style={font=\scriptsize}, ylabel style={font=\scriptsize}, ticklabel style={font=\scriptsize},
  axis x line*=bottom, axis y line*=left, clip=false]
\addplot[black, mark=square*, mark size=1.6pt] coordinates {(50,57.356) (200,171.675) (1000,781.383)};
\addplot[black, dashed, mark=triangle*, mark size=1.8pt] coordinates {(50,17.699) (200,52.545) (1000,238.391)};
\addplot[black, dotted, thick, mark=o, mark size=1.6pt] coordinates {(50,0.27) (200,1.08) (1000,5.4)};
\node[font=\scriptsize, anchor=south east, inner sep=1.5pt] at (axis cs:1000,781) {kernel-per-call, 762{,}440 accesses};
\node[font=\scriptsize, anchor=north west, inner sep=2pt] at (axis cs:200,45) {per-stage, 232{,}400};
\node[font=\scriptsize, anchor=west, inner sep=1.5pt] at (axis cs:52,4.5) {per-signature, 5{,}400 (count $\times$ latency)};
\node[font=\scriptsize, anchor=north west, inner sep=2pt] at (rel axis cs:0.02,1.0) {gem5, processor side only};
\end{axis}
\end{tikzpicture}
\caption{Simulated processor-side cost of one signature.}
\label{fig:gem5}
\end{figure}
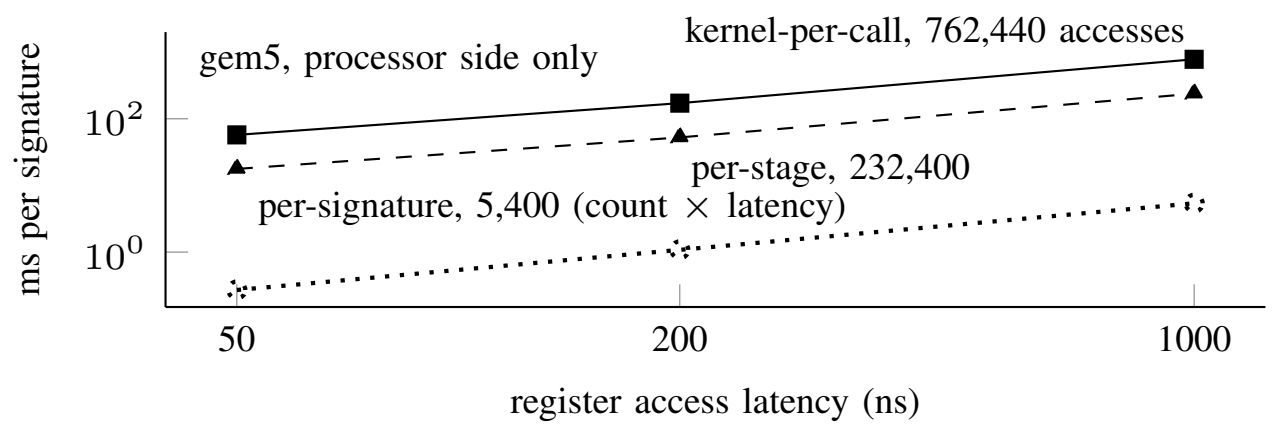

The kernel-per-call figure comes from a transaction-level study of an existing driver for these kernels~\cite{elhadedy2026wepwawet}: replaying that unmodified driver against a register model of the fabric counts the bus accesses it issues, 935{,}635 per mean signature over 252 kernel calls, together with the waits its code schedules.
Priced at the bus-access costs measured on the AUP-ZU3 board (XCZU3EG) through the C driver of this work, 47~ns per write and at most 251~ns per read, the accesses alone take at least 119~ms, and the coded waits add another second.
A gem5~\cite{binkert2011gem5} simulation of the same replay, in syscall-emulation mode with one in-order 1.2~GHz core and a DDR memory model, where the register blocks are uncached stand-ins with a set latency, attributes the processor-side cost to uncached register accesses rather than to DRAM.
Fig.~\ref{fig:gem5} shows how that cost behaves: for one signature of three attempts the simulated time grows with the register access latency, from 57~ms at 50~ns to 781~ms at 1~$\mu$s for per-call control, while the DRAM traffic stays near 222~kB at every granularity.
The count of register accesses times the cost of one is therefore the whole processor-side bill, and only the count moves it by orders of magnitude.
Chaining the kernels of each stage, which the same study models but no hardware implements, removes the redundant loads, initializations, and scrubs and leaves 238{,}959 to 287{,}286 accesses, still 36 to 45~ms of bus time before any wait.
Both are far above Neon software. Only the per-signature granularity, measured on the board in Section~\ref{sec:impl}, brings the total within an order of magnitude of software, with about 5{,}400 bus accesses per signature, a count at which a 1~$\mu$s access costs 5.4~ms of processor time against 781~ms for per-call control, so the design is far less sensitive to the latency of the boundary.

Three design principles follow. The rejection loop, its branch decisions, and its retry bookkeeping must be controlled on chip, so that no kernel call or acceptance check waits for the processor. The operands of an attempt must live in a private, banked on-chip memory that the kernels stream from and to, so that the matrix, the mask, and the intermediate vectors never touch the L2 cache and DRAM that the mission tasks share. And the processor must invoke the fabric once per signature, as one job whose provisioning and readback take the 5{,}400 register accesses counted above, which also makes that boundary the natural place to keep secrets out of the processor's reach.

\section{Proposed Architecture}\label{sec:arch}
\wc realizes the three principles of Section~\ref{sec:analysis} with a ROM microsequencer that owns the rejection loop, a private four-bank scratchpad that the kernels stream from and to, and a host window through which the processor issues one job per signature.

\begin{table}[t]
\caption{\wc resources at 300~MHz.}\label{tab:util}
\centering\footnotesize\setlength{\tabcolsep}{3pt}
\begin{tabular}{lrrrrr}
\toprule
Image / block & LUT & FF & DSP & BRAM & WNS (ns)\\
\midrule
Engine core, Sign + Verify & 44{,}747 & 36{,}637 & 42 & 23\rlap{\,$^a$} & \\
Entropy source & 112 & 195 & 0 & 0 & \\
Mission image, full design & 45{,}410 & 37{,}679 & 42 & 23 & +0.026\\
Sealed, $\rhopp$ in fabric & 44{,}041 & 37{,}038 & 42 & 23 & +0.057\\
Sealed, with KeyGen & 44{,}787 & 37{,}062 & 42 & 23 & +0.011\\
\bottomrule
\end{tabular}\\[2pt]
\raggedright\scriptsize LUT: lookup table; FF: flip-flop; DSP: multiplier slice; BRAM: block RAM tile; WNS: worst negative slack. Post-route on the XCZU3EG: 70{,}560 LUT, 141{,}120 FF, 360 DSP, 216 BRAM tiles; the mission image occupies 90.2\% of the logic slices. $^a$Whole design. Full-design rows include the AXI interconnect. The sealed images also disable configuration readback.
\end{table}

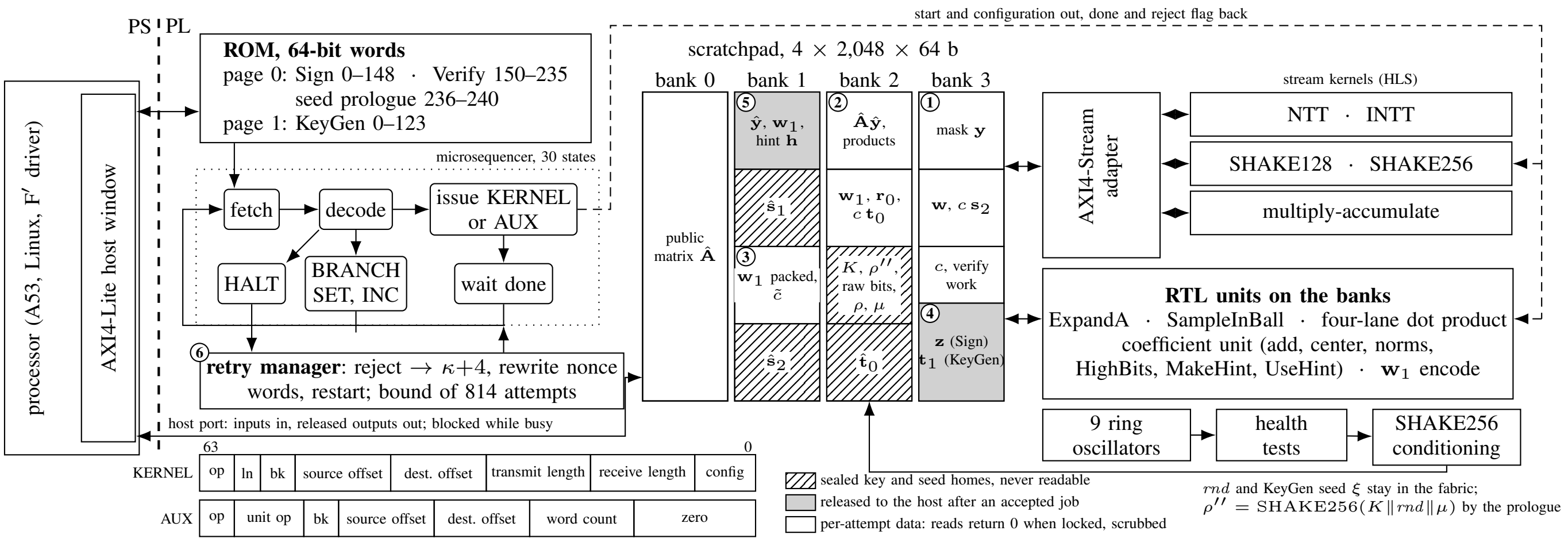
\begin{figure*}[t]
\centering
\begin{tikzpicture}[yscale=0.9, >=Latex, font=\scriptsize,
  bx/.style={draw, align=center, inner sep=2pt, minimum height=5mm},
  st/.style={draw, rounded corners=1.5pt, align=center, inner sep=2pt, minimum height=4.6mm},
  f/.style={draw, minimum height=4.2mm, inner sep=0pt, align=center, font=\tiny},
  key/.style={pattern=north east lines}, rel/.style={fill=black!18},
  num/.style={circle, draw, fill=white, inner sep=0.6pt, font=\tiny\bfseries}]
\node[bx, minimum width=1.55cm, minimum height=4.3cm] (ps) at (0.78,2.2) {};
\node[rotate=90, font=\scriptsize] at (0.33,2.2) {processor (A53, Linux, F$'$ driver)};
\node[bx, minimum width=0.62cm, minimum height=4.0cm] (axi) at (1.2,2.2) {};
\node[rotate=90, font=\scriptsize] at (1.2,2.2) {AXI4-Lite host window};
\draw[dashed, thick] (1.78,-0.12) -- (1.78,5.45);
\node[anchor=north west, font=\scriptsize, inner sep=1.5pt] at (1.8,5.45) {PL};
\node[anchor=north east, font=\scriptsize, inner sep=1.5pt] at (1.76,5.45) {PS};
\node[bx, minimum width=4.5cm, minimum height=1.25cm, align=left, anchor=north west] (rom) at (2.25,5.2)
  {\textbf{ROM, 64-bit words}\\page 0:\ Sign 0--148 \ $\cdot$ \ Verify 150--235\\\phantom{page 0:}\ seed prologue 236--240\\page 1:\ KeyGen 0--123};
\node[st] (fe) at (2.85,2.95) {fetch};
\node[st] (de) at (4.05,2.95) {decode};
\node[st, align=center] (is) at (5.75,2.95) {issue KERNEL\\or AUX};
\node[st] (wa) at (5.75,2.0) {wait done};
\node[st] (br) at (4.05,2.0) {BRANCH\\SET, INC};
\node[st] (ha) at (2.85,2.0) {HALT};
\draw[->] (fe) -- (de); \draw[->] (de) -- (is); \draw[->] (is) -- (wa);
\draw[->] (de) -- (br); \draw[->] (de.south west) -- (ha.north east);
\draw[->] (wa.south) -- ++(0,-0.27) -| ([xshift=-4.7mm]fe.west) -- (fe.west);
\draw (br.south) -- ++(0,-0.27);
\draw[->] ([xshift=-2mm]rom.south -| fe) -- ([xshift=-2mm]fe.north);
\node[draw, dotted, fit=(fe)(de)(is)(wa)(br)(ha), inner xsep=7pt, inner ysep=4.5pt] (seq) {};
\node[anchor=south east, font=\tiny, inner sep=1pt] at (seq.north east) {microsequencer, 30 states};
\node[bx, minimum width=4.5cm, align=center, anchor=north west] (rt) at (2.25,1.12)
  {\textbf{retry manager}: reject $\rightarrow$ $\kappa{+}4$, rewrite nonce\\words, restart; bound of 814 attempts};
\draw[->] (ha.south) -- (ha.south |- rt.north);
\draw[->] (rt.north -| is) -- (seq.south -| is);
\node[num] at (2.25,1.12) {6};
\def\u{0.1}
\newcommand{\fld}[4]{\node[f, minimum width=#2*\u cm] at ({2.25+(#1+#2/2)*\u},#3) {#4};}
\fld{0}{4}{-0.42}{op}\fld{4}{3}{-0.42}{ln}\fld{7}{4}{-0.42}{bk}\fld{11}{11}{-0.42}{source offset}\fld{22}{11}{-0.42}{dest.\ offset}\fld{33}{12}{-0.42}{transmit length}\fld{45}{12}{-0.42}{receive length}\fld{57}{7}{-0.42}{config}
\fld{0}{4}{-1.0}{op}\fld{4}{8}{-1.0}{unit op}\fld{12}{4}{-1.0}{bk}\fld{16}{11}{-1.0}{source offset}\fld{27}{11}{-1.0}{dest.\ offset}\fld{38}{12}{-1.0}{word count}\fld{50}{14}{-1.0}{zero}
\node[anchor=east, font=\tiny, inner sep=2pt] at (2.25,-0.42) {KERNEL};
\node[anchor=east, font=\tiny, inner sep=2pt] at (2.25,-1.0) {AUX};
\node[anchor=south west, font=\tiny, inner sep=1pt] at (2.25,-0.2) {63};
\node[anchor=south east, font=\tiny, inner sep=1pt] at (2.25+64*\u,-0.2) {0};
\def\bw{0.98}\def\bh{3.95}\def\bx{7.35}\def\by{0.5}
\foreach \i/\n in {0/bank 0,1/bank 1,2/bank 2,3/bank 3}{
  \draw ({\bx+\i*(\bw+0.08)},\by) rectangle ++(\bw,\bh);
  \node[font=\scriptsize, anchor=south, inner sep=1.5pt] at ({\bx+\i*(\bw+0.08)+\bw/2},\by+\bh) {\n};
}
\node[font=\scriptsize, anchor=south] at ({\bx+2*\bw+0.12},\by+\bh+0.27) {scratchpad, 4 $\times$ 2{,}048 $\times$ 64 b};
\newcommand{\reg}[5]{
  \pgfmathsetmacro{\ya}{\by+\bh-(#2)/2048*\bh}\pgfmathsetmacro{\yb}{\by+\bh-(#3)/2048*\bh}
  \path[draw, #4] ({\bx+#1*(\bw+0.08)},\ya) rectangle ({\bx+#1*(\bw+0.08)+\bw},\yb);
  \node[font=\tiny, align=center, fill=white, inner sep=0.8pt] at ({\bx+#1*(\bw+0.08)+\bw/2},{(\ya+\yb)/2}) {#5};}
\newcommand{\regr}[5]{
  \pgfmathsetmacro{\ya}{\by+\bh-(#2)/2048*\bh}\pgfmathsetmacro{\yb}{\by+\bh-(#3)/2048*\bh}
  \path[draw, #4] ({\bx+#1*(\bw+0.08)},\ya) rectangle ({\bx+#1*(\bw+0.08)+\bw},\yb);
  \node[font=\tiny, align=center, inner sep=0.8pt] at ({\bx+#1*(\bw+0.08)+\bw/2},{(\ya+\yb)/2}) {#5};}
\reg{0}{0}{2048}{}{public\\matrix $\hat{\mathbf A}$}
\regr{1}{0}{512}{rel}{$\hat{\mathbf y}$, $\mathbf w_1$,\\hint $\mathbf h$}
\reg{1}{512}{1024}{key}{$\hat{\mathbf s}_1$}
\reg{1}{1024}{1536}{}{$\mathbf w_1$ packed,\\$\tilde c$}
\reg{1}{1536}{2048}{key}{$\hat{\mathbf s}_2$}
\reg{2}{0}{512}{}{$\hat{\mathbf A}\hat{\mathbf y}$,\\products}
\reg{2}{512}{1024}{}{$\mathbf w_1$, $\mathbf r_0$,\\$c\,\mathbf t_0$}
\reg{2}{1024}{1536}{key}{$K$, $\rho''$,\\raw bits,\\$\rho$, $\mu$}
\reg{2}{1536}{2048}{key}{$\hat{\mathbf t}_0$}
\reg{3}{0}{512}{}{mask $\mathbf y$}
\reg{3}{512}{1024}{}{$\mathbf w$, $c\,\mathbf s_2$}
\reg{3}{1024}{1400}{}{$c$, verify\\work}
\regr{3}{1400}{2048}{rel}{$\mathbf z$ (Sign)\\$\mathbf t_1$ (KeyGen)}
\path[draw, key] (9.0,-0.62) rectangle ++(0.34,0.2); \node[anchor=west, font=\tiny, inner sep=1.5pt] at (9.34,-0.52) {sealed key and seed homes, never readable};
\path[draw, rel] (9.0,-0.9) rectangle ++(0.34,0.2); \node[anchor=west, font=\tiny, inner sep=1.5pt] at (9.34,-0.8) {released to the host after an accepted job};
\draw (9.0,-1.18) rectangle ++(0.34,0.2); \node[anchor=west, font=\tiny, inner sep=1.5pt] at (9.34,-1.08) {per-attempt data: reads return 0 when locked, scrubbed};
\draw[<->] (axi.east |- 0,4.2) -- (2.25,4.2);
\draw[<->] (axi.east |- 0,0.05) -- (7.15,0.05) -- (7.15,0.8) -- (\bx,0.8);
\node[font=\tiny, anchor=south west, inner sep=1pt] at (1.85,0.06) {host port: inputs in, released outputs out; blocked while busy};
\node[bx, minimum width=1.35cm, minimum height=1.9cm, anchor=north west] (ad) at (11.95,4.45) {};
\node[rotate=90, font=\scriptsize, align=center] at (12.62,3.5) {AXI4-Stream\\adapter};
\node[bx, minimum width=3.7cm, anchor=north west, align=center] (k1) at (13.65,4.45) {NTT \ $\cdot$ \ INTT};
\node[bx, minimum width=3.7cm, anchor=north west, align=center] (k2) at (13.65,3.82) {SHAKE128 \ $\cdot$ \ SHAKE256};
\node[bx, minimum width=3.7cm, anchor=north west, align=center] (k3) at (13.65,3.19) {multiply-accumulate};
\node[anchor=south, font=\tiny, inner sep=1pt] at (15.5,4.47) {stream kernels (HLS)};
\foreach \k in {k1,k2,k3}{\draw[<->] (ad.east |- \k) -- (\k.west);}
\draw[<->] ({\bx+4*\bw+0.24},3.5) -- (ad.west |- 0,3.5);
\node[bx, minimum width=5.4cm, minimum height=1.5cm, anchor=north west, align=center] (aux) at (11.95,2.2)
  {\textbf{RTL units on the banks}\\ExpandA \ $\cdot$ \ SampleInBall \ $\cdot$ \ four-lane dot product\\coefficient unit (add, center, norms,\\HighBits, MakeHint, UseHint) \ $\cdot$ \ $\mathbf w_1$ encode};
\draw[<->] ({\bx+4*\bw+0.24},1.55) -- (aux.west |- 0,1.55);
\draw[densely dashed] (is.east) -- (6.98,2.95) -- (6.98,5.3) -- (17.7,5.3) -- (17.7,1.55);
\draw[->, densely dashed] (17.7,0 |- k2.east) -- (k2.east);
\draw[->, densely dashed] (17.7,1.55) -- (aux.east |- 0,1.55);
\node[font=\tiny, anchor=south, inner sep=1pt] at (12.4,5.3) {start and configuration out, done and reject flag back};
\node[bx, anchor=north west, minimum width=1.7cm] (ro) at (11.95,0.4) {9 ring\\oscillators};
\node[bx, anchor=north west, minimum width=1.5cm] (ht) at (13.95,0.4) {health\\tests};
\node[bx, anchor=north west, minimum width=1.7cm] (cd) at (15.75,0.4) {SHAKE256\\conditioning};
\draw[->] (ro) -- (ht); \draw[->] (ht) -- (cd);
\draw[->] (cd.south) -- ++(0,-0.08) -| ({\bx+2*(\bw+0.08)+\bw/2},\by);
\node[font=\tiny, anchor=north west, inner sep=1pt, align=left] at (13.75,-0.5) {$\mathit{rnd}$ and KeyGen seed $\xi$ stay in the fabric;\\$\rho''=\mathrm{SHAKE256}(K\Vert\mathit{rnd}\Vert\mu)$ by the prologue};
\node[num] at ({\bx+3*(\bw+0.08)+0.13},{\by+\bh-0.13}) {1};
\node[num] at ({\bx+2*(\bw+0.08)+0.13},{\by+\bh-0.13}) {2};
\node[num] at ({\bx+1*(\bw+0.08)+0.13},{\by+\bh-1024/2048*\bh-0.13}) {3};
\node[num] at ({\bx+3*(\bw+0.08)+0.13},{\by+\bh-1400/2048*\bh-0.13}) {4};
\node[num] at ({\bx+1*(\bw+0.08)+0.13},{\by+\bh-0.13}) {5};
\end{tikzpicture}
\caption{The \wc engine: Sign, Verify, KeyGen, and the sealed key regions.}
\label{fig:fsm}
\end{figure*}

The microsequencer comes first, because it is what replaces the processor. Each 64-bit ROM word carries a 4-bit opcode in bits 63:60, and its remaining fields are laid out per opcode so that one fetch fully specifies an operation, as Fig.~\ref{fig:fsm} shows.
A KERNEL word names a kernel lane, its source and destination banks and offsets, the stream lengths, and an entry in a table of kernel configurations; an AUX word names one of the register-transfer-level (RTL) units with the same bank and offset fields. The other opcodes move words to the configuration path, update the attempt counter $\kappa$, branch on the sticky reject flag or the attempt limit, and halt.

The sequencer is a 30-state finite-state machine (FSM).
A fetch is followed by decode; a KERNEL word then loads its configuration through a local AXI4-Lite master, arms the receive and transmit streams, starts the kernel, and waits for its completion acknowledgment, while an AUX word starts its unit and waits for done. Every waiting state is guarded by a watchdog, and every error path passes through a scrub of the scratchpad before the error is reported. The ROM holds three programs (Table~\ref{tab:prog}), and a mode register selects the entry address. The engine therefore turns from signing to verification or key generation by starting the next job at another address: no datapath reconfiguration is required, and each job begins at its ROM entry address when the engine is free. Verification (Algorithm~\ref{alg:verify}) is a single pass over the same units with the peer's public key and no secret, at a cost of 86 ROM words and no second engine. The sealed variants add a five-word prologue that forms $\rhopp$ in the fabric.

\begin{table}[t]
\caption{The three \wc programs.}\label{tab:prog}
\centering\footnotesize\setlength{\tabcolsep}{3pt}
\begin{tabular}{lrrrrrcl}
\toprule
Program & Words & S128 & S256 & NTT & INTT & Loop & Key\\
\midrule
Sign$^a$ & 149 & 16 & 6 & 5 & 16 & yes & reads sealed\\
Verify & 86 & 16 & 2 & 5 & 8 & no & public only\\
KeyGen & 124 & 16 & 19 & 16 & 4 & no & writes sealed\\
\bottomrule
\end{tabular}\\[2pt]
\raggedright\scriptsize ROM words and kernel calls per job; S128 and S256 are SHAKE-128 and SHAKE-256, INTT the inverse NTT. $^a$Per attempt.
\end{table}

Fig.~\ref{fig:fsm} lets one signing attempt be followed by eye along the circled numbers. SHAKE-256 expands $\rhopp$ and $\kappa$ into the mask in bank~3; the transforms and the dot-product unit form $\hat{\mathbf{A}}\hat{\mathbf{y}}$ in bank~2 against the matrix in bank~0; the packed high bits give the challenge in bank~1; the response is built in bank~3 from a key half that never leaves bank~1; the hint lands at the top of bank~1; and a reject at any check hands control to the retry manager.

The programs act on operands that never leave the chip. They live in four banks of 2{,}048 64-bit words.
Four banks let a stage read its two operands and write its result in different banks, so the kernels stream at one word per cycle without port conflicts; $\hat{\mathbf{A}}$ fills one bank, and the 11-bit offsets address any word of a bank directly. A stream adapter turns a bank range into an AXI-Stream (AXIS) burst toward the selected kernel and writes the returned stream into the destination bank, so the kernels, written for high-level synthesis (HLS), see only streams and never the processor's interconnect. The kernels are the same family of HLS forward and inverse NTT, SHAKE-128, SHAKE-256, and multiply-accumulate units used in~\cite{elhadedy2026wepwawet};
the remaining steps of Algorithm~\ref{alg:sign} are RTL units on the auxiliary bus, among them the samplers, a four-lane dot-product unit (dot4), and a coefficient unit for centering, addition, rounding, hints, and the norm checks. The host reaches each bank through a second port behind an AXI4-Lite window; while a program runs, host accesses to the scratchpad are answered with a slave error, so the processor cannot observe or disturb a job in flight.

With control and memory on chip, the rejection loop closes without the processor. A failed norm check sets a sticky reject flag; the program branches to an increment of $\kappa$ by four and halts the attempt, and a retry manager restores the 40 scratchpad words that hold the mask-seed and nonce state and the seed $\rho$, in 87 to 90 cycles, and relaunches the program without the processor, up to the FIPS~204 bound of 814 attempts.
The hint-weight bound $\|\mathbf{h}\|_1\le\omega=80$ is the only acceptance condition that counts rather than compares.
A 10-bit counter beside the coefficient unit accumulates the ones written by the four MakeHint passes and raises the reject flag before the accept branch, at no cycle cost.

The loop draws its randomness from the same fabric. Hedged signing needs 32 fresh bytes $rnd$ per signature from inside the chip. Following the XOR-of-rings construction analyzed in~\cite{sunar2007trng}, nine rings of three inverters are XORed, synchronized into the 300~MHz domain through two flip-flops, sampled at a programmable divider, checked by a windowed health monitor, and packed into words.
Each signature consumes 1{,}024 raw bits, screened by the repetition-count and adaptive-proportion tests of NIST SP~800-90B~\cite{sp80090b} for an assumed 0.5 bit of min-entropy per sample and conditioned with SHAKE-256.
An assessment of a one-million-bit capture that assumes no independence between samples estimated 0.743 bit per raw sample, about 760 bits for each 256-bit output.
In the mission image the tests and the conditioning run in the driver; the sealed image moves both into the fabric, together with the derivation $\rhopp=H(K\|rnd\|\mu)$ (Section~\ref{sec:security}). The source is not a validated SP~800-90B module: restart tests, characterization across temperature and voltage corners, and the clock-domain-crossing assertions on the ring outputs remain open.

These pieces give the fabric a simple cost model. With $C_a$ cycles per attempt, $N$ attempts, a retry overhead of $C_r\approx 90$ cycles, $C_v$ cycles per verification, and clock frequency $f$, the fabric signs in $T_s=(N C_a+(N-1)C_r)/f$ and verifies in a single pass $T_v=C_v/f$.
Because $C_a$ is the same for every attempt to within a few tens of cycles in simulation, the fabric time of a signature is its attempt count scaled by $C_a/f$, independent of caches, interrupts, or processor load.

\section{Circuit Evaluation}\label{sec:impl}
The second question has two parts: whether the complete control fits on a small device at a useful clock, and what then limits it relative to dedicated cores and software. \wc was implemented with Vivado and Vitis HLS 2022.2 for the XCZU3EG of the AUP-ZU3 board, and every image in Table~\ref{tab:util} meets timing at 300~MHz after routing with no routing or design-rule errors.
The board is a commercial part with no radiation tolerance and serves as a prototype of the architecture; a flight version would add configuration scrubbing and protection of the stored key.
The engine uses 64\% of the LUTs but 90\% of the logic slices, so the device is placement-limited rather than logic-limited. The sealed images, which add the in-fabric seed derivation, the access control, the scrub, and key generation, change the LUT count by less than the variation between placement runs, because they reuse the same datapath.

Both programs were checked against the reference implementation in simulation and on the board. Six deterministic signing vectors, including acceptance after 20 attempts and at a hint weight of exactly 80, reproduce the reference signatures and attempt counts byte for byte, and verification accepts the valid signature while rejecting a modified message, a modified challenge, and a $\mathbf{z}$ outside its norm bound.
On each board a standalone run of the sealed key-generation image generated its key in the fabric, 325{,}829 and 325{,}942 cycles, and produced 2{,}000 hedged signatures that all verify under PQClean against the fabric's public key, with every host sweep of the scratchpad clean.

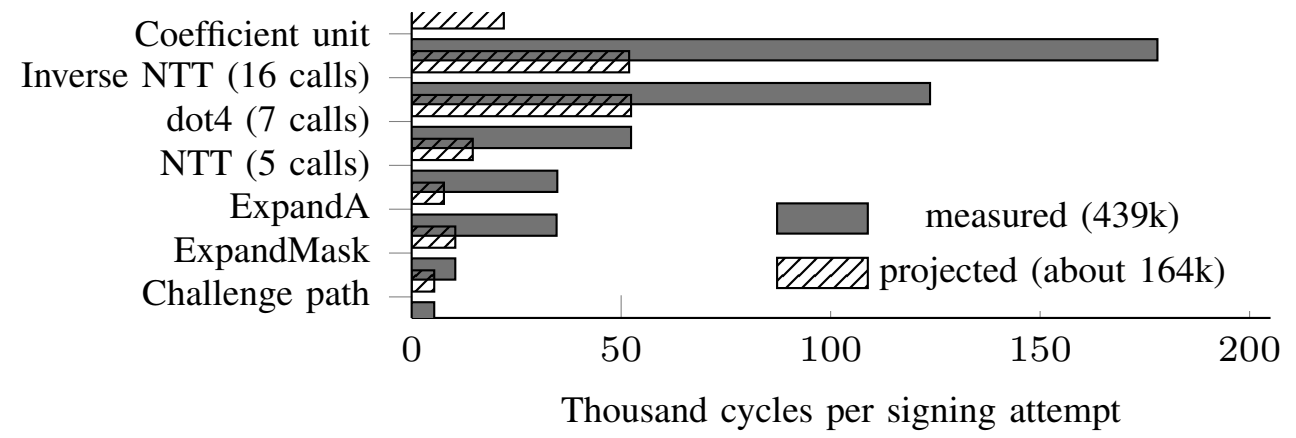
\begin{figure}[t]
\centering
\begin{tikzpicture}
\begin{axis}[xbar, width=0.82\columnwidth, height=3.6cm, bar width=4pt, xmin=0, xmax=205,
  xlabel={Thousand cycles per signing attempt}, xlabel style={font=\scriptsize},
  ytick={1,...,7}, yticklabels={Challenge path, ExpandMask, ExpandA, NTT (5 calls), dot4 (7 calls), Inverse NTT (16 calls), Coefficient unit},
  yticklabel style={font=\scriptsize}, xticklabel style={font=\scriptsize}, enlarge y limits=0.08,
  axis x line*=bottom, axis y line*=left, area legend, legend style={font=\scriptsize, draw=none, at={(0.98,0.02)}, anchor=south east}]
\addplot[draw=black, fill=black!55] coordinates {(5.388,1) (10.416,2) (34.592,3) (34.765,4) (52.366,5) (123.728,6) (178.000,7)};
\addplot[draw=black, pattern=north east lines] coordinates {(5.388,1) (10.416,2) (7.704,3) (14.587,4) (52.366,5) (51.906,6) (22.000,7)};
\legend{measured (439k), projected (about 164k)}
\end{axis}
\end{tikzpicture}
\caption{Where the cycles of one signing attempt go.}
\label{fig:profile}
\end{figure}

Correctness established, the question becomes where the cycles go. In the profile one signing attempt takes 439{,}270 cycles, 1.464~ms at 300~MHz, and verification 214{,}186 cycles; with the measured mean of 4.49 attempts a signature costs the fabric 6.57~ms.
Fig.~\ref{fig:profile} attributes the attempt to its causes from a cycle-accurate trace of every ROM instruction. Data movement is not among them: streaming between the scratchpad and the kernels costs 4.6\% of the attempt and kernel configuration 0.3\%, which confirms the analysis of Section~\ref{sec:analysis}.
Three quarters of the attempt come from two implementation choices instead. The coefficient unit processes its 7{,}168 words per attempt with a serial state machine at about 25 cycles per word. The HLS transforms spend most of each call starting and finishing an inner pipelined loop once per butterfly group, so only about a third of their cycles execute butterflies.
Table~\ref{tab:units} sets those cycles against area. The units that account for 88.5\% of the attempt, the coefficient unit, the transforms, and the multiply-accumulate, occupy about 5{,}000 LUTs, 11\% of the design, whereas the two SHAKE kernels alone hold 61\% of the LUTs and, with the samplers, serve 11.5\% of the cycles; speed can therefore be bought with little area, and area would be recovered in the hash kernels.

\begin{table}[t]
\caption{Cycles and resources by unit.}\label{tab:units}
\centering\footnotesize\setlength{\tabcolsep}{3pt}
\begin{tabular}{lrrrrr}
\toprule
Unit & Cycles (\%) & LUT & FF & DSP & BRAM\\
\midrule
Coefficient unit & 40.5 & 2{,}091 & 749 & 8 & 0\\
Inverse transform & 28.2 & 900 & 1{,}134 & 9 & 2.5\\
Forward transform & 7.9 & 507 & 847 & 7 & 2.5\\
Multiply-accumulate & 11.9 & 1{,}543 & 1{,}201 & 18 & 2\\
SHAKE kernels, samplers & 11.5 & 31{,}781 & 28{,}891 & 0 & 0\\
Control and memory$^a$ & -- & 6{,}413 & 2{,}425 & 0 & 16\\
\bottomrule
\end{tabular}\\[2pt]
\raggedright\scriptsize Mission image. $^a$Sequencer, scratchpad, stream adapter, retry manager, configuration; their 5.2\% of the cycles is counted in the rows they serve. BRAM in tiles. Rows are hierarchical post-route counts and do not sum exactly to Table~\ref{tab:util}.
\end{table}

The profile ranks three changes by risk. Expanding $\hat{\mathbf{A}}$ once per signature instead of once per attempt is a ROM change, since the matrix is never overwritten, at the price of holding it across attempts, where a radiation upset would persist rather than be rewritten within one attempt. Pipelining the coefficient unit to one word every three cycles and flattening the transforms' butterfly loop, as the fully pipelined butterfly units of KiD do~\cite{mandal2024kid}, are circuit changes that use the free DSP and block-RAM capacity. Together they project to about 0.74 million cycles, 2.5~ms, per mean signature, a projection from measured counts rather than a measurement, which still leaves the fabric about 2.2 times slower than Neon.

Against that profile, Table~\ref{tab:compare} places \wc beside published Dilithium and ML-DSA hardware at the same security level. The dedicated cores are 12 to 62 times faster per average signature at comparable LUT counts, and their area-time products are 22 to 51 times better; Neon software on the same chip is 6.0 times faster for signing and 1.9 times for verification.
What that speed is traded for is in Table~\ref{tab:feat}. \wc spends it on generality, with Sign, Verify, and KeyGen as programs over one set of units rather than one pipeline per operation, and its 6.6~ms is close to the roughly 7~ms of Sapphire's Dilithium-II signing at 72~MHz~\cite{banerjee2019sapphire}, the other program-driven design with a measured average; HORCRUX, which keeps the flow in software, needs 1.8 times the cycles of \wc.

\begin{table*}[t]
\caption{Comparison with prior ML-DSA-44 hardware.}\label{tab:compare}
\centering\footnotesize\setlength{\tabcolsep}{5pt}
\begin{tabular}{llrrrrrrrcrr}
\toprule
Design & Device & MHz & LUT & FF & DSP & BRAM & Sign (kcyc.) & Sign (ms) & Basis & Verify (ms) & kLUT$\cdot$ms\\
\midrule
Beckwith~\cite{beckwith2021} & Virtex US+ & 256 & 53{,}907 & 28{,}435 & 16 & 29 & 29.9 & 0.117 & E & 0.026 & 6.3\\
Land~\cite{land2021}$^a$ & Artix-7 & 140 & 24{,}320 & 9{,}668 & 45 & 15 & 76.6 & 0.547 & M & 0.141 & 13.3\\
KaLi~\cite{aikata2023kali}$^b$ & Zynq US+ & 270 & 23{,}277 & 9{,}758 & 4 & 24 & 31.7 & 0.117 & S & 0.057 & n.c.\\
ML-DSA-OSH~\cite{norga2026mldsaosh} & US+$^c$ & 230 & 54{,}942 & 29{,}309 & 16 & 29 & 24.3 & 0.106 & A & 0.029 & 5.8\\
HORCRUX~\cite{dolmeta2025horcrux}$^e$ & Zynq US+ & 15 & 2{,}820 & 395 & 0 & 0 & 3{,}579 & 238.6 & A & 124.2 & n.c.\\
PQClean Neon~\cite{becker2022neon,kannwischer2022pqclean} & A53 & 1{,}200 & -- & -- & -- & -- & -- & 1.099 & M & 0.371 & --\\
\textbf{This work} & XCZU3EG & 300 & 44{,}859$^d$ & 36{,}832$^d$ & 42 & 23 & 1{,}970.9 & 6.570 & M & 0.714 & 295\\
\bottomrule
\end{tabular}\\[2pt]
\raggedright\scriptsize
Basis of the signing figures: M, measured mean over the rejection distribution (this work, 4{,}234 signatures with 4.49 attempts on average; Neon, 1{,}000 runs on the AUP-ZU3); E, best case scaled by 4.25 expected attempts; S, single attempt; A, average as reported, method not stated. $^a$Authors' revised version: mean of 1{,}000 simulated runs including pre-computation, times derived at 140~MHz; \cite{beckwith2021} and \cite{norga2026mldsaosh} list 163~MHz and 27{,}433 LUTs for the same cycle counts. $^b$Unified Kyber/Dilithium processor; sum of its three reported signing phases. $^c$Device family as reported. $^d$Signer core plus entropy source, excluding interconnect; the full design is in Table~\ref{tab:util}. $^e$Coprocessor beside a RISC-V core, flow in software; average of 100 simulated runs, times derived at 15~MHz. kcyc.: thousand cycles. US+: UltraScale+. n.c.: not comparable. kLUT$\cdot$ms uses the signing time.
\end{table*}

\section{Integration with Flight Software}\label{sec:system}
The third question, what the engine gives and costs inside flight software, starts at the interface. F$'$~\cite{bocchino2018fprime} composes flight software from components that communicate through typed ports. On each node (Fig.~\ref{fig:node}) \texttt{DilithiumSign} and \texttt{DilithiumVerify} expose the two operations as commands with telemetry, \texttt{PqcTransport} carries frames to the peer, and \texttt{PqcSecurityGateway} enforces a persistent sequence-number replay window; the four are pinned to the four cores. A passive \texttt{HwAccelMgr} component, invoked on the caller's thread, links a C driver and is the only code that touches the fabric.

The interface is deliberately narrow. The driver maps the two AXI4-Lite windows, writes the operands into the scratchpad, launches the program, and waits by sleeping in 100~$\mu$s steps while polling a status bit, so the calling core returns to the scheduler while the fabric works; there is neither direct memory access nor an interrupt line.
This processor-side work is dominated by reading back and packing the signature, so it is fixed per signature while the fabric time varies with the attempt count; Section~\ref{sec:mission} measures both.
Verification is fail-closed: if the fabric image is absent, lacks the verification program, or reports an error, \texttt{HwAccelMgr} rejects the frame and raises a high-severity event, and it never falls back to software silently.
A software signer remains linked only as an explicitly selected backend for the comparison of Section~\ref{sec:mission}.

\section{Evaluation Under Mission Load}\label{sec:mission}
The third question closes with a measurement: what the engine does to the flight software it serves, under real mission functions rather than synthetic loads. Every fabric run uses the sealed key-generation image with fabric-generated keys certified as in Section~\ref{sec:security}.

The ADCS task closes the loop around a simulated 3U CubeSat in low Earth orbit: in each period it propagates the true attitude with ten integration substeps, runs a multiplicative extended Kalman filter (MEKF)~\cite{markley2014} on simulated measurements, and applies a proportional-derivative (PD) quaternion feedback law, at 100~Hz as a \texttt{SCHED\_FIFO} thread of priority 80 on core~1 with the period as its deadline.
The payload compresses imager frames with the CCSDS~121.0 lossless algorithm~\cite{ccsds121} through libaec~\cite{libaec}, on $1024\times1024$ 16-bit tiles cut from a Sentinel-2 Level-1C scene~\cite{drusch2012s2}.
Its rate follows from a real imager: a MultiScape100-class push-broom camera at 500~km~\cite[Table~3]{multiscape_sensors23} yields 85.0~MB/s of 16-bit samples during a pass, so three payload threads on cores 3, 0, and 2 each compress 13.5 tiles per second with one tile period as deadline.

The command rate is a parameter, set at 10 signed frames per second, at which the fabric is busy 6.5\% of the time, so the experiment tests the signer's tail and its sharing of cores and caches rather than its average cost.
Four configurations run on node~1 under the same workload: no signing, fabric signing, Neon, and Neon with the flight software at real-time priority 70. Each is repeated three times in alternating order, with signing measured over a 60~s window inside a 75~s mission load, on Linux 6.6 without real-time patches. Two-node runs then have both nodes sign and verify, 10 frames per second each way, under the same load on both.

\begin{table*}[t]\caption{\wc and Neon under the mission workload.}\label{tab:mission}\centering\footnotesize\setlength{\tabcolsep}{3.4pt}
\begin{tabular}{lccccccc}\toprule
Configuration & Signed of 600 & Sign median (ms) & 99\% within (ms) & Core time$^a$ (ms) & ADCS delay$^b$ (ms) & ADCS misses & Payload finish (s)\\\midrule
No signing & -- & -- & -- & -- & 0.41--0.44 & 0 & 75.0\\
Fabric & 594/584/597 & 5.12--5.13 & 21.2--33.0 & 0.53 / 0.60 & 0.52--0.85 & 0 & 75.0--76.9\\
Neon & 596/597/594 & 0.91--1.04 & 3.2--4.0 & 0.94 / 3.9 & 0.47--1.18 & 0 & 75.0--75.3\\
Neon, priority 70 & 596/593/593 & 0.93--0.94 & 3.5--3.8 & 0.93 / 3.8 & 0.29--0.53 & 0 & 75.0--75.6\\
\midrule
Two nodes, node~1 & 597/599/595 & 5.13--5.15 & 25.6 & 0.53 / 0.60 & 0.59--1.08 & 0 & 75.0--79.1\\
Two nodes, node~2 & 596/597/591 & 5.11--5.16 & 27.0--31.5 & 0.53 / 0.59 & 0.30--0.72 & 0 & 75.1--75.5\\
\bottomrule\end{tabular}\\[2pt]\raggedright\scriptsize Fabric rows use the sealed image with certified keys and Neon rows a software key; 10 signed commands per second, three runs per configuration; ranges are across runs. Rows 1 to 4 are node~1 signing alone; in the last two rows each node also verifies the frames of its peer. Signing is measured over a 60~s window inside 75~s of ADCS and payload load; the payload finish time is the wall time to compress the 75~s of imagery. $^a$Time the calling core is occupied per signature, median / 99\% within. $^b$Worst release delay of the 100~Hz task.
\end{table*}

The engine leaves the mission undisturbed. In every run of Table~\ref{tab:mission} the ADCS task completed each of its 7{,}500 control periods before the next release, its worst release delay stayed below 1.2~ms of the 10~ms period, and the payload finished its 75~s of imagery within 4.2~s of the end of the pass.
At this command rate Neon does not disturb it either, so the two signers differ in what a signature costs the processor and in where the key lives. A fabric signature occupies the calling core for 0.53~ms at the median and at most 0.60~ms for 99\% of the 5{,}350 fabric signatures, whatever the number of rejections, because a retry never returns to the processor; the core is released while the fabric works.
A Neon signature occupies its core for 0.94~ms at the median, 3.9~ms at 99\%, and up to 5.4~ms, and its rejection loop runs on the caches the mission tasks share.

The fabric's own timing does not depend on the load. A linear fit over the 5{,}350 signatures gives $C_a=439{,}276$ cycles per attempt, six cycles from the profile of Section~\ref{sec:impl}, plus 35{,}345 cycles per signature for the entropy draw, the seed derivation, and the scrub; every signature lies within 306 cycles, 1~$\mu$s, of that line while all four cores are 76 to 89\% busy.
The service-time tail in Table~\ref{tab:mission} is therefore the attempt-count distribution of Section~\ref{sec:prelim}, here 4.33 attempts on average, and nothing else.

Between two nodes the authenticated link holds under full load. The fabrics produced 1{,}791 and 1{,}784 signatures with no signing failure and no flight-software assertion; each receiver verified every frame of its peer in its own fabric, in 214{,}161 to 214{,}197 cycles and 1.6~ms at the median through F$'$, accepted all 3{,}575, and rejected none.
What remains on the processor is almost entirely the readback and packing of the signature through register reads; packing it in the fabric and returning it by direct memory access (DMA) with a completion interrupt would, by an estimate, bound the processor-side work at about 0.27~ms.

\section{Security Boundary and Key Lifecycle}\label{sec:security}
The fourth question is whether the one-job boundary that Section~\ref{sec:analysis} chose for cost also keeps the secrets, and the sealed images make it do so. The signer must keep $\mathbf{s}_1$, $\mathbf{s}_2$, $\mathbf{t}_0$, $K$, and every $\rhopp$ confidential, because any of them, or the mask $\mathbf{y}$ of a single attempt, rejected or not, yields a forgery; the verifier needs no secret but must hold the authentic public key of its peer; and a processor under an attacker's control can still ask an intact signer to sign anything. Three properties are at stake, and they are established to different degrees. Isolation of the secrets from the processor bus is enforced by the fabric and measured on hardware. Authenticated distribution of public keys is demonstrated with ground-signed certificates that both nodes check, the ground public key being a file on each node. Hardware-rooted identity and remote attestation are not established: a compromised processor can replace the bitstream or the ground key file, and a remote verifier has no evidence that a peer key originated in a trusted fabric configuration.

The sealed images keep the key and all secret-derived values in the fabric. Once the core is locked, host reads of key storage and of every intermediate return zero and increment a sticky violation counter, writes to key storage are discarded, the scratchpad is scrubbed after every job, on request, and on reset, and the bitstream disables configuration readback.
The entropy draw, its health tests, the SHAKE-256 conditioning, and $\rhopp=H(K\|rnd\|\mu)$ run in the fabric, so the processor never sees $rnd$ or $\rhopp$. In the key-generation image a third ROM program derives the key pair from a fabric seed $\xi$ and the host can read only the public key.
On the AUP-ZU3, with a host-provisioned test key, the same message signed twice gave two different valid signatures; full sweeps of the 8{,}192-word window after signing and after zeroization found no secret word, and an independent root attack program found nothing.
With no key file anywhere, the fabric generated a key and signed under it (Section~\ref{sec:impl}), refused key generation while locked, and on a second generation produced a different key whose signatures did not verify under the first.
Simulation extends these results to every attempt count, retry exhaustion, entropy faults, and resets during signing and key generation, against a trace of every secret value.
The readback setting was not exercised, and side channels, fault injection, and corner characterization of the entropy source are not evaluated.

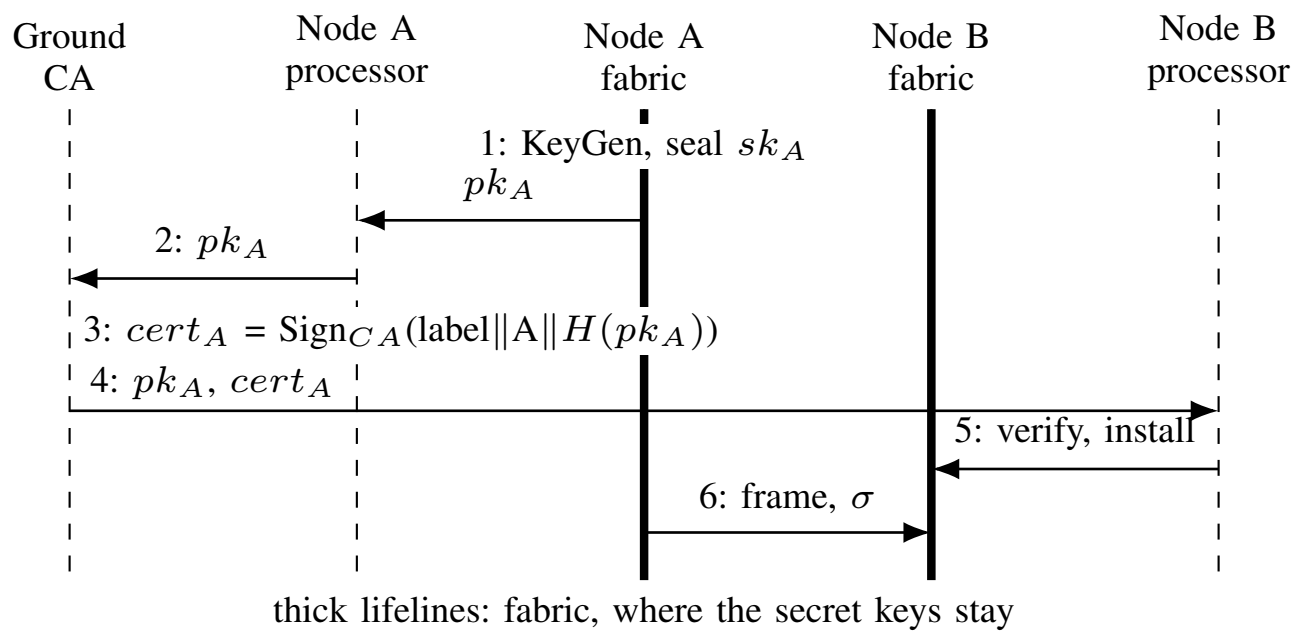
\begin{figure}[t]
\centering
\begin{tikzpicture}[yscale=0.7, font=\scriptsize, >=Latex, m/.style={->, line width=0.5pt}]
\foreach \x/\n in {0/{Ground\\CA}, 1.9/{Node A\\processor}, 3.8/{Node A\\fabric}, 5.7/{Node B\\fabric}, 7.6/{Node B\\processor}}
  { \node[align=center, anchor=south] at (\x,0) {\n}; \draw[dashed] (\x,0) -- (\x,-4.45); }
\draw[line width=1.6pt] (3.8,0) -- (3.8,-4.45); \draw[line width=1.6pt] (5.7,0) -- (5.7,-4.45);
\node[fill=white, inner sep=1pt] at (3.8,-0.35) {1: KeyGen, seal $sk_A$};
\draw[m] (3.8,-1.05) -- node[above, pos=0.5]{$pk_A$} (1.9,-1.05);
\draw[m] (1.9,-1.6) -- node[above]{2: $pk_A$} (0,-1.6);
\node[fill=white, inner sep=1pt, anchor=west] at (0.05,-2.1) {3: $cert_A$ = Sign$_{CA}$(label$\|$A$\|H(pk_A)$)};
\draw[m] (0,-2.85) -- node[above, pos=0.125, fill=white, inner sep=1pt, yshift=1pt]{4: $pk_A$, $cert_A$} (7.6,-2.85);
\draw[m] (7.6,-3.4) -- node[above]{5: verify, install} (5.7,-3.4);
\draw[m] (3.8,-4.0) -- node[above]{6: frame, $\sigma$} (5.7,-4.0);
\node[anchor=north, inner sep=2pt] at (3.8,-4.5) {thick lifelines: fabric, where the secret keys stay};
\end{tikzpicture}
\caption{The \wc key certification protocol.}
\label{fig:proto}
\end{figure}

A peer must still learn which public key belongs to which node, so the sealed engine is paired with the six-step certification protocol of Fig.~\ref{fig:proto}, in which $sk_A$, $pk_A$, and $cert_A$ are the secret key, public key, and certificate of node~A. Each fabric generates its key pair from on-chip entropy and releases only the public key, which is downlinked. The ground station, acting as certificate authority (CA), signs a labeled statement binding the node name to a SHAKE-256 digest of that key with its own ML-DSA-44 key and uplinks the peer's key and certificate. A node installs a peer key only after verifying its certificate against the ground public key provisioned before launch, and crosslink frames are then signed in the sender's fabric with sender, destination, and sequence number inside the signed envelope.

The steps differ in maturity. Generation and sealing (step 1) and fabric signing and verification between two sealed, certified nodes under load (step 6, Section~\ref{sec:mission}) are measured on hardware. Steps 2 to 5 were exercised with the two fabric-generated keys: each node checked its peer's ground-signed certificate with the fabric Verify program and with PQClean, accepted it, and rejected copies with one bit of the key or of the certificate altered.
In a negative test, a receiver left with a stale peer key rejected all 1{,}219 frames it received and accepted none.
In this demonstration the ground public key is a file on the node, which root can replace. Anchoring it as a constant of the bitstream, with the verdict decided and $\mu$ formed in the fabric, passes in simulation only, and it would tie the verdict to the loaded bitstream without attesting that bitstream to a remote party.

The protocol answers the threats in turn. A substituted public key fails the certificate check as long as the ground key file is intact, since only the ground key can bind a node name to a key; the certificate records that the ground accepted a key for that name, not that the key was generated in a sealed fabric. A replayed frame fails the sequence check of the signed envelope. Theft of the storage medium yields nothing, because no secret is ever written to it. A read of the signer from the processor returns zeros, as measured. What remains is what root can do around the boundary: request signatures of its own messages, which a policy gate holding the sequence number and a rate limit in the fabric would restrict but not eliminate; replace the bitstream or the ground key file, against which the defence is authenticated boot~\cite{ug908}; and reset the node, after which the volatile key must be regenerated and recertified. ML-DSA-87, which CNSA~2.0 requires~\cite{cnsa2}, needs new constants and programs and an estimated 3.5 times the matrix storage.

\section{Conclusion}
The cost of a hardware-assisted lattice signature is set by how often the processor crosses into the fabric, and the control-granularity analysis shows that one job per signature is the only boundary that takes the processor off the critical path. \wc builds that boundary on a small FPGA: a ROM microsequencer, a banked scratchpad, and a retry manager run Sign, Verify, and key generation as three programs over one datapath, and the same boundary seals the key, the entropy, and the mask seed from the processor. Inside flight software the engine costs the processor a fixed half millisecond per signature whatever the number of rejections, its own timing is unaffected by a saturating workload, and two sealed nodes authenticated every frame they exchanged without a missed attitude deadline. The engine trades the throughput of a dedicated pipeline for that generality and isolation, and its profile places the difference in two serial units. Node identity is not yet rooted in hardware. The next steps are in-fabric signature packing with DMA readback, a bitstream-anchored trust chain, a radiation-tolerant implementation, and ML-DSA-87 programs.

\section*{Acknowledgment}
This work was supported by AFRL (FA8650-24-2-2403), the U.S. Department of Defense (W911NF-24-1-0265), the U.S. Navy NEEC (N00174-23-1-0002), and the ONR Summer Faculty Research Program; the views expressed are the authors' own. Generative AI tools (e.g., ChatGPT) assisted with technical editing, supporting scripts, and figure preparation based on the authors' designs. The research architecture, scientific methodology, experimental validation, and conclusions were conceived, directed, and evaluated by the authors.

\raggedbottom
\bibliographystyle{IEEEtran}
\bibliography{refs}
\vspace{-3.2\baselineskip}
\begin{IEEEbiography}[{\includegraphics[width=1in,height=1.25in,clip,keepaspectratio]{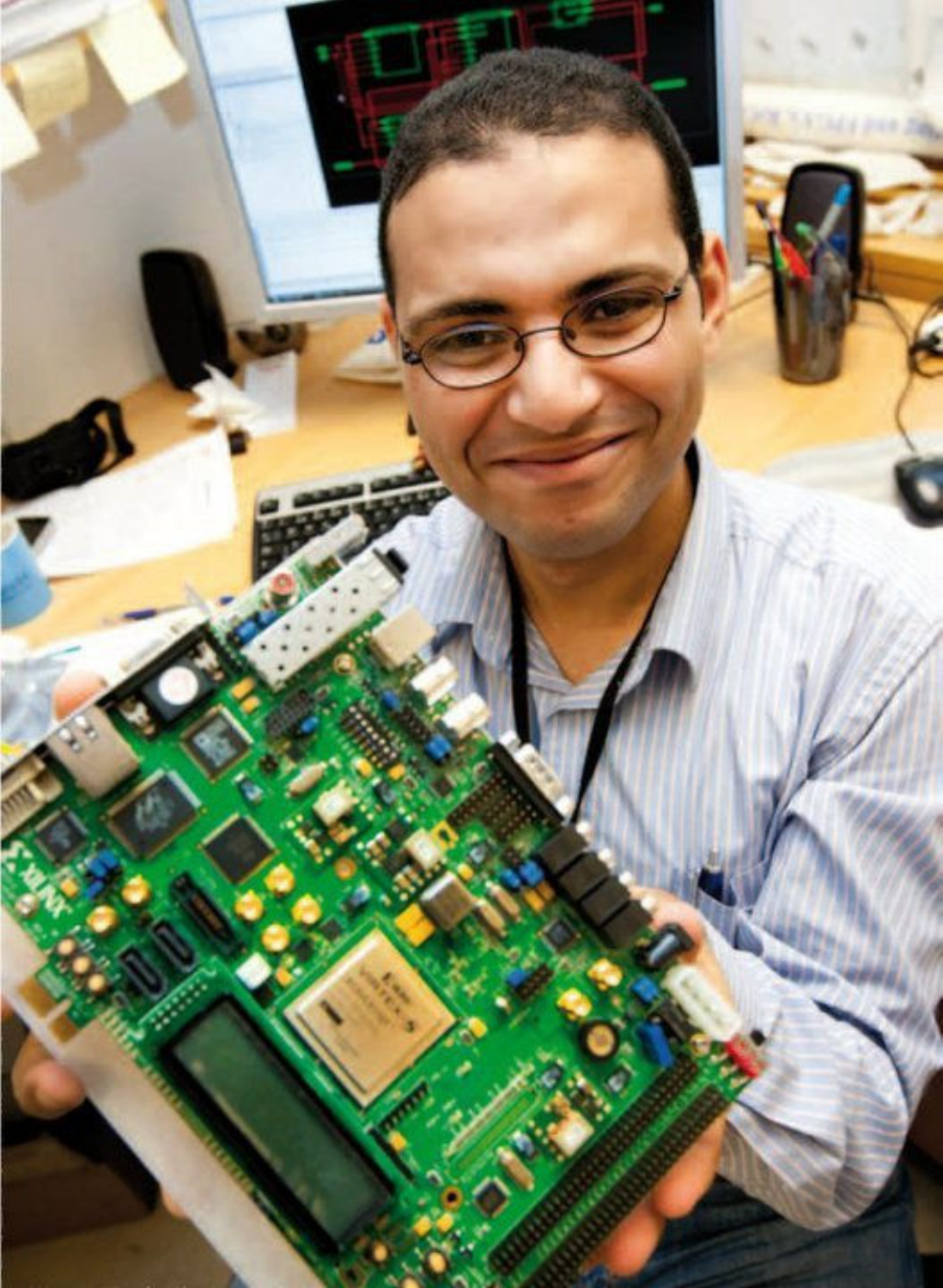}}]{Mohamed El-Hadedy}
(Senior Member, IEEE) is a Professor of Electrical and Computer Engineering at California State Polytechnic University, Pomona, where he directs the Reconfigurable Space Computing Laboratory, and a Senior Research Scientist with the Coordinated Science Laboratory, University of Illinois at Urbana-Champaign. He received the Ph.D. degree from the Norwegian University of Science and Technology. His research covers reconfigurable hardware security for space and autonomous systems, from RTL to flight-software integration.
\end{IEEEbiography}
\vspace{-3.2\baselineskip}
\begin{IEEEbiography}[{\includegraphics[width=1in,height=1.25in,clip,keepaspectratio]{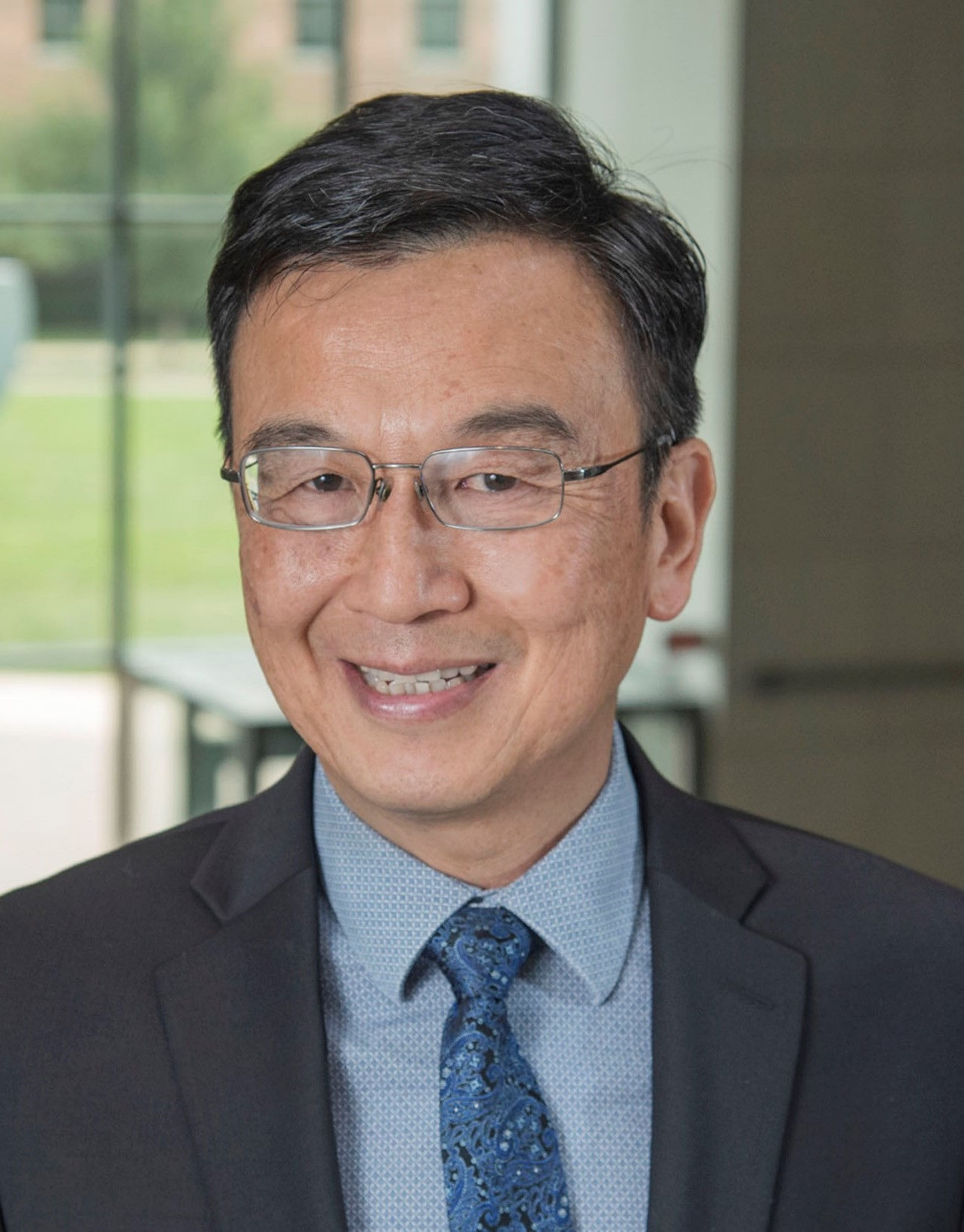}}]{Wen-Mei Hwu}
(Fellow, IEEE) is a Senior Distinguished Research Scientist at NVIDIA and Professor Emeritus at the University of Illinois at Urbana-Champaign, where he spent more than three decades as a Professor, Sanders-AMD Endowed Chair, and Chief Scientist of the Parallel Computing Institute. In 2008 he became director of NVIDIA's first CUDA Center of Excellence, and in 2010 he co-authored the textbook \emph{Programming Massively Parallel Processors: A Hands-on Approach}. His research covers parallel architecture, compilers, and infrastructure software for data-intensive and AI applications. He received the 2024 ACM-IEEE CS Eckert-Mauchly Award.
\end{IEEEbiography}

\end{document}